\documentclass[aps,showpacs,preprintnumbers,amsmath,amssymb]{revtex4}
\usepackage{dcolumn}
\usepackage[dvips]{epsfig}

\begin{document}
\baselineskip=0.8 cm
\title{\bf Thin accretion disk around a Kaluza-Klein black
hole with squashed horizons}

\author{Songbai Chen\footnote{csb3752@163.com}, Jiliang Jing
\footnote{jljing@hunnu.edu.cn}}

\affiliation{Institute of Physics and Department of Physics, Hunan
Normal University,  Changsha, Hunan 410081, People's Republic of
China \\ Key Laboratory of Low Dimensional Quantum Structures \\
and Quantum Control of Ministry of Education, Hunan Normal
University, Changsha, Hunan 410081, People's Republic of China}

\begin{abstract}
\baselineskip=0.6 cm
\begin{center}
{\bf Abstract}
\end{center}

We study the accretion process in the thin disk around a squashed
Kaluza-Klein black hole and probe the effects of the extra
dimensional scale $\rho_0$ on the physical properties of the disk.
Our results show that with the increase of the parameter $\rho_0$,
the energy flux, the conversion efficiency, the radiation
temperature, the spectra luminosity and the spectra cut-off
frequency of the thin accretion disk decrease, but the inner border
of the disk increases. This implies that the extra dimension scale
imprints in the mass accretion process in the disk.

\end{abstract}

 \maketitle
\newpage
\section{Introduction}

String theory is widely believed to be the most promising candidate
theory for a unified description of everything in our Universe,
which predicts the existence of the extra dimension. Thus, the
detection of the extra dimension has attracted a great deal of
attention recently since it can present the signature of the string
and the modification of string theory.  It is shown that the
quasinormal modes originated from the perturbation around high
dimensional black holes \cite{Shen, Abdalla,Chen,Kanti} carry the
peculiar information about the extra dimension, which would be
tested in the gravitational wave probe in the near future. Moreover,
the spectrum of Hawking radiation from the high dimensional black
holes could provide another possible way to observe the extra
dimension, which is expected to be detected in particle accelerator
experiments \cite{5, 6, 7, 8, 9, 10, 11, 12, 13}. The investigations
of the strong gravitational lensing of the brane world black holes
\cite{bgl} indicate that the extra dimension imprints in the
deflection angle, the angular position and magnification of the
relativistic images, which implies that the extra dimension might be
observed by measuring these lensing parameters in the astronomical
experiments.

Since the accretion processes is a powerful indicator of the
physical nature of the central celestial objects, the analysis of
the signatures of the accretion disk  could provide another possible
way to detect the extra dimension. The accretion disk is such a
structure formed by the diffuse material in orbital motion around a
central compact body, which now is an important research topic in
the astrophysics. The simplest theoretical model of the accretion
disks is the steady-state thin accretion disk model in which the
disk has negligible thickness \cite{sdk1,sdk2,Page,Thorne}. In this
model, the disk is in hydrodynamical equilibrium since the heat
generated by stress and dynamic friction in the disk can be
dispersed through the radiation over its surface. Moreover, the mass
accretion rate in the disk maintains a constant and is independent
of time variable. The physical properties of matter forming a thin
accretion disk in a variety of background spacetimes have been
investigated extensively in
\cite{Harko,Bhattacharyya,Kovacs,Torres,Yuan,Guzman,Pun}. The
special signatures appeared in the energy flux and the emission
spectrum emitted by the disk can provide us not only the information
about black holes in the Universe, but also the profound
verification of alternative theories of gravity. Therefore, the
study of properties of the thin disk around high dimensional black
holes could help us to probe the extra dimension in astronomical
observation in the future.

The Kaluza-Klein black hole with squashed horizons is a new kind of
interesting Kaluza-Klein type metrics
\cite{IM,sq2,sq3,sq4,TVN,sq5,sq6}, which is the solution of the
Einstein-Maxwell equations in the five-dimensional spacetime.  This
family of black holes have a structure of the five-dimensional black
holes in the vicinity of the black hole horizon, but are locally the
direct product of the four-dimensional Minkowski spacetime and the
circle in the region far from the black holes. Recently, the Hawking
radiation has been considered in these squashed Kaluza-Klein black
holes under the low-energy approximation, which indicates that the
luminosity of Hawking radiation can tell us the size of the extra
dimension which could open a window to detect extra dimensions
\cite{sq1hw,sq2hw}. The quasinormal modes for the scalar and
gravitational perturbations in the background of the Kaluza-Klein
black hole with squashed horizons have been investigated in
\cite{sqq1,sqq2}, which implies that the quasinormal frequencies
contain the information of the extra dimension. Moreover, the
precession of a gyroscope in a circular orbit in the squashed
Kaluza-Klein black hole spacetime have been studied in \cite{Ksq},
which shows that the modification from the extra dimension is
proportional to the square of the ratio between the size of extra
dimension and  the gravitational radius of central object. The study
of the strong gravitational lensing of a squashed Kaluza-Klein black
hole also manifests that the size of the extra dimension imprints in
the observational lensing parameters, such as deflection angle, the
angular position and magnification of the relativistic images
\cite{schen}. The main purpose of the present Letter is to study the
properties of the thin accretion disk in the squashed Kaluza-Klein
black hole spacetime and see whether it can leave us the signature
of the extra dimension in the energy flux and the emission spectrum
emitted in the mass accretion process.

The Letter is organized as follows: in the following section we will
present the geodesic equations for the timelike particles moving in
the equatorial plane in the squashed Kaluza-Klein black hole
spacetime. In Sec.III, we study the physical properties of the thin
accretion disk around the squashed Kaluza-Klein black hole and probe
the effects of the extra dimension on the energy flux, temperature,
emission spectrum and efficiency of the thin accretion disks onto
this black hole. We end the paper with a summary.

\section{The geodesic equations in the squashed Kaluza-Klein black
hole spacetime}

The five-dimensional neutral static Kaluza-Klein black hole with
squashed horizons is described by \cite{IM,sq4}
\begin{eqnarray}
ds^2=-f(\rho)dt^2+\frac{K}{f(\rho)}d\rho^2+K\rho^2(d\theta^2+\sin^2\theta
d\phi^2)+\frac{r^2_{\infty}}{4K}(d\psi+\cos^2\theta d\phi)^2,
\label{metric0}
\end{eqnarray}
with
\begin{eqnarray}
f(\rho)=1-\frac{\rho_H}{\rho},\;\;\;\;\;\;\;\;\;\;K=1+\frac{\rho_0}{\rho},
\end{eqnarray}
where the angular coordinates $\theta\in [0,\pi)$,  $\phi\in
[0,2\pi)$, and $\psi\in [0,4\pi)$. $\rho_H$ is the radius of the
black hole event horizon. $r_{\infty}$ corresponds to the spatial
infinity. The quantity $\rho_0$ is related to the parameters
$r_{\infty}$ and $M$ by $\rho^2_{0}=\frac{r^2_{\infty}-M}{4}$. The
Komar mass of the squashed Kaluza-Klein black hole can be described
by $M=\pi r_{\infty}\rho_H/G_5$ \cite{Ksq,Ksm}, where $G_5$ is the
five-dimensional gravitational constant. Moreover, one can find that
the event horizon radius $\rho_H$ can be written further as
$\rho_H=2G_4M$ since the relationship between $G_5$ and $G_4$ ( the
four-dimensional gravitational constant) can be expressed as
$G_5=2\pi r_{\infty}G_4$ \cite{Ksq,Ksm} in this black hole
spacetime. Since the metric (\ref{metric0}) reduces to a
four-dimensional Schwarzschild black hole with a constant twisted
$S_1$ fiber as $\rho_0\rightarrow 0$ and tends to a spherical
symmetrical five-dimensional Schwarzschild black hole as
$\rho_H\ll\rho_0$, $\rho_0$ can be regarded as a scale of transition
from five-dimensional spacetime to an effective four-dimensional
one.

In order to probe the properties of the thin accretion disk in
squashed Kaluza-Klein black hole spacetime (\ref{metric0}), we must
study the geodesics equations of motion for the particle moving in
this neutral static background. As in
Refs.\cite{Page,Thorne,Harko,Bhattacharyya,Kovacs,Torres,Yuan,Guzman,Pun,Ksq,schen},
we here consider only the orbits in the equatorial plane. The
$\theta$-component of the geodesics for the particle moving in the
orbits with the condition $\theta=\pi/2$ tells us \cite{schen}
\begin{eqnarray}
\frac{d\phi}{d\lambda}\frac{d\psi}{d\lambda}=0,
\end{eqnarray}
where $\lambda$ is an affine parameter along the geodesics. This
means that either $\frac{d\phi}{d\lambda}=0$ or
$\frac{d\psi}{d\lambda}=0$. Here we set $\frac{d\psi}{d\lambda}=0$,
so that we can compare with the results obtained in the
four-dimensional black hole spacetime. In doing so, one can find
that the timelike geodesics equations can be simplified as
\begin{eqnarray}
&&\frac{dt}{d\lambda}=\frac{\tilde{E}}{g_{tt}}=\frac{\tilde{E}}{f(\rho)},\nonumber\\
&&\frac{d\phi}{d\lambda}=\frac{\tilde{L}}{g_{\phi\phi}}=\frac{\tilde{L}}{K\rho^2},\nonumber\\
&&g_{tt}g_{\rho\rho}\bigg(\frac{d\rho}{d\lambda}\bigg)^2+V_{eff}(\rho)=\tilde{E}^2,
\end{eqnarray}
where $\tilde{E}$ and $\tilde{L}$ are the specific energy and the
specific angular momentum of the particle, respectively. The
effective potential $V_{eff}(\rho)$ has the form
\begin{eqnarray}
V_{eff}(\rho)=g_{tt}\bigg(1+\frac{\tilde{L}^2}{g_{\phi\phi}}\bigg)
=\bigg(1-\frac{\rho_H}{\rho}\bigg)\bigg[1+\frac{\tilde{L}^2}{\rho(\rho+\rho_0)}\bigg].
\end{eqnarray}
For stable circular orbits in the equatorial plane, we have
$V_{eff}(\rho)=\tilde{E}^2$ and $V_{eff}(\rho)_{,\rho}=0$. Making
use of these conditions, one can get the specific energy
$\tilde{E}$, the specific angular momentum $\tilde{L}$, and the
angular velocity $\Omega$ of the particle moving in circular orbit
in the squashed Kaluza-Klein black hole spacetime
\begin{eqnarray}
&&\tilde{E}=\frac{g_{tt}}{\sqrt{g_{tt}-g_{\phi\phi}\Omega^2}}
=\sqrt{\frac{(2\rho+\rho_0)(\rho-\rho_H)^2}{\rho^2(2\rho-3\rho_H)+\rho\rho_0(\rho-\rho_H)}}\;,\label{sE}\\
&&\tilde{L}=\frac{g_{\phi\phi}\Omega}{\sqrt{g_{tt}-g_{\phi\phi}\Omega^2}}
=\sqrt{\frac{\rho(\rho+\rho_0)^2\rho_H}{\rho(2\rho-3\rho_H)+\rho_0(\rho-\rho_H)}}\;, \label{sL}\\
&&\Omega=\frac{d\phi}{dt}=\sqrt{\frac{g_{tt,\rho}}{g_{\phi\phi,\rho}}}=\sqrt{\frac{\rho_H}{\rho^2(2\rho+\rho_0)}}.
\end{eqnarray}
The marginally stable orbit of the particle around the black hole is
given by the condition $V_{eff}(\rho)_{,\rho\rho}=0$. Combining
Eqs.(\ref{sE}), (\ref{sL}) with $V_{eff}(\rho)_{,\rho\rho}=0$, one
can find that the radius $\rho_{ms}$ of the marginally stable orbit
for the particle moving in the squashed Kaluza-Klein black hole
spacetime can be expressed as
\begin{eqnarray}
\rho_{ms}=\rho_H\bigg[1+\bigg(1+\frac{\rho_0}{\rho_H}\bigg)^{\frac{1}{3}}+\bigg(1+\frac{\rho_0}{\rho_H}\bigg)^{\frac{2}{3}}\bigg].\label{pms}
\end{eqnarray}
With the increase of $\rho_0$, one can find that for the particle
moving in the circular orbit, the specific energy $\tilde{E}$, the
specific angular momentum $\tilde{L}$, and the marginally stable
orbit radius $\rho_{ms}$ increase, but the angular velocity $\Omega$
decreases.  As the parameter $\rho_0\rightarrow 0$, the above
quantities $\tilde{E}$,  $\tilde{L}$, $\Omega$, and $\rho_{ms}$ are
reduced to those in the four-dimensional Schwarzschild black hole
spacetime.

\section{The properties of thin accretion disks in the squashed Kaluza-Klein black
hole spacetime}

In this section, we will investigate the accretion process in the
thin disk around the squashed Kaluza-Klein black hole and probe the
effects of the extra dimensional scale $\rho_0$ on the physical
properties of the thin accretion disk. For the thin accretion disk
model in the squashed Kaluza-Klein black hole spacetime, we can
assume similarly that the central plane of the disk is located in
the equatorial plane of the black hole and the disk height $H$ (the
maximum half thickness of the disk) is much smaller than the
characteristic radius $\rho$ of the disk, i.e., $H\ll \rho$.
Moreover, the thin disk can stay in hydrodynamical equilibrium since
the heat generated by stress and dynamic friction in the disk can be
dispersed through the radiation over its surface, which makes the
disk stabilize its thin vertical size. For simplification, we here
assume that the thin disk around the squashed Kaluza-Klein black
hole is modeled by a steady state accretion disk \cite{Page,Thorne}
in which the mass accretion rate $\dot{M_0}$ stays a constant. As in
the disk around the usual black holes
\cite{Page,Thorne,Harko,Bhattacharyya,Kovacs,Torres,Yuan,Guzman,Pun},
we can also measure the physical quantities describing the accreting
matter by averaging over a characteristic time scale $\Delta t$, the
azimuthal angle $\Delta\phi=2\pi$, and the height $H$ of the disk.

Similarly, we suppose that the accreting matter in the disk can be
described by an anisotropic fluid with the energy-momentum tensor
\cite{Page,Thorne}
\begin{eqnarray}
T^{\mu\nu}=\varepsilon_0
u^{\mu}u^{\nu}+2u^{(\mu}q^{\nu)}+t^{\mu\nu},
\end{eqnarray}
where the rest mass density $\varepsilon_0$, the energy flow vector
$q^{\mu}$ and the stress tensor $t^{\mu\nu}$ of the accreting matter
are defined in the averaged rest-frame of the orbiting particle with
four-velocity $u^{\mu}$. In the averaged rest-frame, we have
$u_{\mu}q^{\mu}=0$ and $u_{\mu}t^{\mu\nu}=0$ since both $q^{\mu}$
and $t^{\mu\nu}$ is orthogonal to $u^{\mu}$ \cite{Page,Thorne}. With
the laws of conservation of the rest mass, the energy and the
angular momentum, one can obtain three time-averaged radial
structure equations of the thin disk around the squashed
Kaluza-Klein black hole
\begin{eqnarray}
&&\dot{M_0}=-2\pi(\rho+\rho_0)\Sigma(\rho) u^{\rho}=\text{Const},\\
&&[\dot{M_0}\tilde{E}-2\pi
\Omega(\rho+\rho_0)W_{\phi}^{\;\rho}]_{,\rho}=2\pi
(\rho+\rho_0)F(\rho)\tilde{E},\label{Ws1}\\
&&[\dot{M_0}\tilde{L}-2\pi
(\rho+\rho_0)W_{\phi}^{\;\rho}]_{,\rho}=2\pi
(\rho+\rho_0)F(\rho)\tilde{L},\label{Ws2}
\end{eqnarray}
where  a dot represents the derivative with respect to the time
coordinate $t$ \cite{Page,Thorne}. The averaged rest mass density
$\Sigma(\rho)$ and the averaged torque $W_{\phi}^{\;\rho}$ are given
by
\begin{eqnarray}
\Sigma(\rho)=\int^{H}_{-H}\langle\varepsilon_0\rangle
\;dz,\;\;\;\;\;\;\; W_{\phi}^{\;\rho}=\int^{H}_{-H}\langle
t_{\phi}^{\;\rho} \rangle \;dz,
\end{eqnarray}
respectively. The quantity $\langle t_{\phi}^{\;\rho}\rangle$ is the
average value of the $\phi-\rho$ component of the stress tensor over
a characteristic time scale $\Delta t$ and the azimuthal angle
$\Delta\phi=2\pi$. Combining Eqs.(\ref{Ws1}) and (\ref{Ws2}) with
the energy-angular momentum relation for circular geodesic orbits
$\tilde{E}_{,\rho}=\Omega \tilde{L}_{,\rho}$, one can eliminate
$W_{\phi}^{\;\rho}$ and obtain the expression of the energy flux
\begin{eqnarray}
F(\rho)=-\frac{\dot{M_0}}{4\pi(\rho+\rho_0)}
\frac{\Omega_{,\rho}}{(\tilde{E}-\Omega\tilde{L})^2}\int^{\rho}_{\rho_{ms}}
(\tilde{E}-\Omega\tilde{L})\Omega_{,\rho}d\rho.\label{enf}
\end{eqnarray}
As in Refs.\cite{Harko}, we here consider the mass accretion driven
by black holes with a total mass of $M=10^6M_{\odot}$, and with a
mass accretion rate of $\dot{M_0}=10^{-12}M_{\odot}\;yr^{-1}$ . In
Fig. (1), we present the total energy flux $F(\rho)$ radiated by a
thin disk around the squashed Kaluza-Klein black hole for different
$\rho_0$. It is shown that the energy flux $F(\rho)$ decreases with
the extra dimensional scale $\rho_0$. The main mathematical reason
is that the presence of $\rho_0$ increases the marginally stable
orbit radius $\rho_{ms}$ and enhances the lower limit of integral in
the energy flux (\ref{enf}). The position of the peak value of
$F(\rho_0)$ moves along the right with the increase of $\rho_0$.
\begin{figure}[ht]
\begin{center}
\includegraphics[width=8cm]{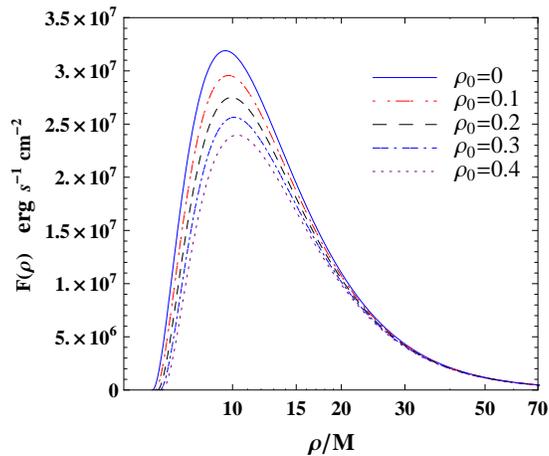}
\caption{Variety of the energy flux $F(\rho)$ with the parameter
$\rho_0$ in the thin disk around the squashed Kaluza-Klein black
hole. Here, we set the total mass of the black hole
$M=10^6M_{\odot}$ and the mass accretion rate
$\dot{M_0}=10^{-12}M_{\odot}\;yr^{-1}$.}
\end{center}
\end{figure}

Let us now to probe the effect of $\rho_0$ on the conversion
efficiency $\epsilon$ for the thin accretion disk around the
squashed Kaluza-Klein black hole. The conversion efficiency is
another important characteristics of the mass accretion process,
which describes the capability of the central object converting rest
mass into outgoing radiation. In general, the conversion efficiency
can be given by the ratio of two rates measured at infinity
\cite{sdk2,Page}: the rate of the radiation energy of photons
escaping from the disk surface to infinity and the mass-energy
transfer rate of the central compact object in the mass accretion.
If all the emitted photons can escape to infinity, one can find that
the efficiency $\epsilon$ is related to the specific energy measured
at the marginally stable orbit $\rho_{ms}$ by
\begin{eqnarray}
\epsilon=1-\tilde{E}_{ms}.\label{effi}
\end{eqnarray}
Substituting Eqs.(\ref{sE}) and (\ref{pms}) into Eq. (\ref{effi}),
we can probe the effects of $\rho_0$ on the conversion efficiency
$\epsilon$ for the mass accretion process occurred in the squashed
Kaluza-Klein black hole spacetime. The dependence of $\epsilon$ on
$\rho_0$ is plotted in Fig.(2), which shows that the larger values
of $\rho_0$ leads to a much smaller efficiency $\epsilon$. This
means that the conversion efficiency of the thin accretion disk in
the squashed Kaluza-Klein black hole spacetime is less than that in
the four-dimensional Schwarzschild one.
\begin{figure}[ht]
\begin{center}
\includegraphics[width=8cm]{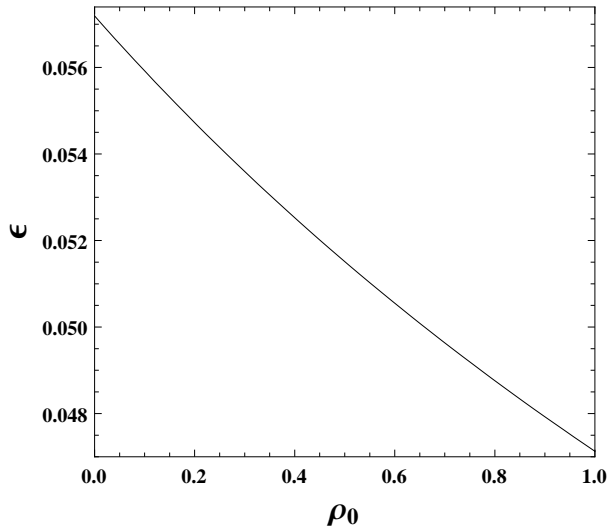}
\caption{Variety of the efficient $\epsilon$ with the parameter
$\rho_0$ for the thin disk around the squashed Kaluza-Klein black
hole.}
\end{center}
\end{figure}

In the steady-state thin disk model \cite{Page,Thorne}, the
accreting matter is generally assumed to be in thermodynamical
equilibrium. This means that the radiation emitted by the disk
surface can be considered as a perfect black body radiation. The
radiation temperature $T(\rho)$ of the disk is related to the energy
flux $F(\rho)$ through the expression
$T(\rho)=[F(\rho)/\sigma]^{1/4}$, where $\sigma$ is the
Stefan-Boltzmann constant, respectively. Thus, one can find that the
dependence of $T(\rho)$ on $\rho_0$ is similar to that of the energy
flux $F(\rho)$ on $\rho_0$, which is also shown in Fig.(3). As in
\cite{Torres}, the observed luminosity $L(\nu)$ for the thin
accretion disk around the squashed Kaluza-Klein black hole can be
expressed as
\begin{eqnarray}
L(\nu)=4\pi
d^2I(\nu)=\frac{16\pi^2h\cos{\gamma}}{c^2}\int^{\rho_f}_{\rho_{i}}\frac{\nu^3\rho
d\rho}{e^{h\nu/KT(\rho)}-1},\label{emspe}
\end{eqnarray}
\begin{figure}[ht]
\begin{center}
\includegraphics[width=8cm]{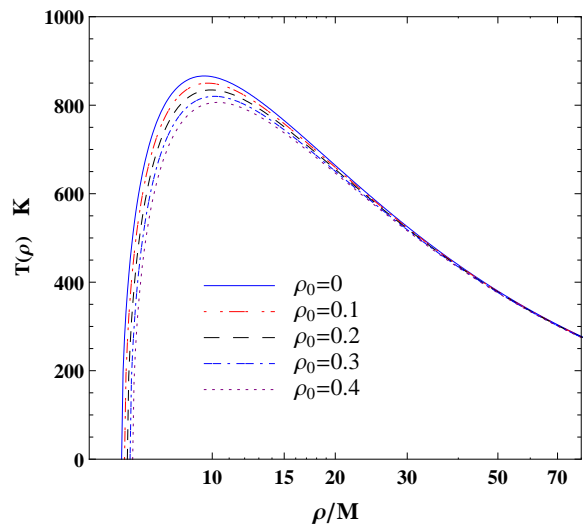}
\caption{Variety of the temperature $T$ with the parameter $\rho_0$
for the thin disk around the squashed Kaluza-Klein black hole. Here,
we set the total mass of the black hole $M=10^6M_{\odot}$ and the
mass accretion rate $\dot{M_0}=10^{-12}M_{\odot}\;yr^{-1}$.}
\end{center}
\end{figure}
\begin{figure}[ht]
\begin{center}
\includegraphics[width=8cm]{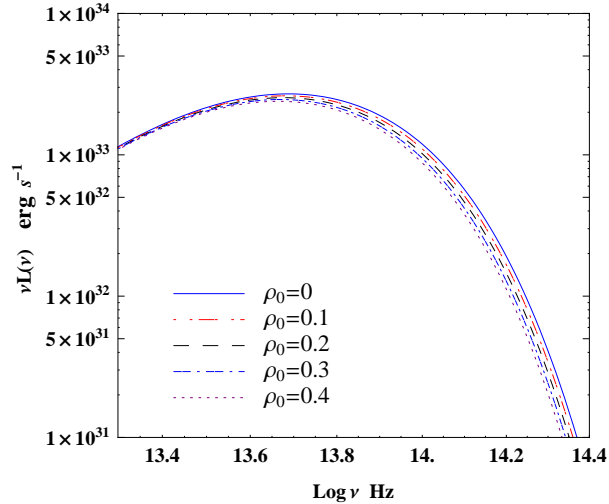}
\caption{Variety of the emission spectrum with the parameter
$\rho_0$ for the thin disk around the squashed Kaluza-Klein black
hole. Here, we set the total mass of the black hole
$M=10^6M_{\odot}$, the mass accretion rate
$\dot{M_0}=10^{-12}M_{\odot}\;yr^{-1}$, and the disk inclination
angle $\gamma=0^{\circ}$, respectively.}
\end{center}
\end{figure}
where $d$ is the distance to the source, $I(\nu)$ is the thermal
energy flux radiated by the disk, and  $\gamma$ is the disk
inclination angle. The quantities $\rho_f$ and $\rho_i$ are the
outer and inner border of the disk, respectively. In order to
calculate the luminosity $L(\nu)$ of the disk, we choose
$\rho_i=\rho_{ms}$ and $\rho_f\rightarrow \infty$ since the flux
over the disk surface vanishes at $\rho_f\rightarrow \infty$ in the
squashed Kaluza-Klein black hole spacetime. Applying the formulas
(\ref{emspe}),  we present the spectral energy distribution of the
disk radiation in Fig.(4). It is shown that the observed luminosity
of the disk decreases with the extra dimensional scale $\rho_0$ in
the total frequency domain ranges. The effect of $\rho_0$ becomes
more distinct as $\nu>5\times10^{13}$ Hz for the chosen values of
$\dot{M}_0$, $M$ and $\gamma$. Moreover, we also find that the
larger value of $\rho_0$ leads to the lower cut-off frequencies,
which means that the spectra of the disk becomes softer in the
squashed Kaluza-Klein black hole background.

\section{summary}

In summary, we have studied the properties of the thin accretion
disk around the squashed Kaluza-Klein black hole and found that the
size of the extra dimension imprints in the energy flux, temperature
distribution and emission spectra of the disk. With the increase of
the parameter $\rho_0$, except the inner border of disk, the energy
flux, the conversion efficiency, the radiation temperature, the
spectra luminosity and cut-off frequency of the thin accretion disk
decrease. The presence of the lower cut-off frequencies means that
the spectra of the disk becomes softer in the squashed Kaluza-Klein
black hole background. Theoretically, we could detect the effects of
the extra dimension on the accretion around the black hole by the
astronomical observations and then make a constraint on the
parameter $\rho_0$. However, the sensitivity of the current
observations is far from being sufficient to detect these effects.
Perhaps with the development of technology, the effects of the extra
dimension on the accretion may be detected in the future.

\section{\bf Acknowledgments}

This work was  partially supported by the National Natural Science
Foundation of China under Grant No.10875041, the NCET under Grant
No.10-0165, the PCSIRT under Grant No. IRT0964 and the construct
program of key disciplines in Hunan Province. J. Jing's work was
partially supported by the National Natural Science Foundation of
China under Grant Nos. 10875040 and 10935013; 973 Program Grant No.
2010CB833004.

\end{document}